\documentclass[letterpaper, 10 pt, conference]{ieeeconf}  % Comment this line out if you need a4paper

\IEEEoverridecommandlockouts                              % This command is only needed if 
\usepackage[maxbibnames=4]{biblatex}
\usepackage{amsmath,amssymb,amsfonts}
\usepackage{algorithm}
\usepackage{algpseudocode}
\usepackage{graphicx}
\usepackage{subfigure}
\usepackage{textcomp}
\usepackage{xcolor}
\title{A General Auxiliary Controller for Multi-agent Flocking}

\author{Jinfan Zhou, Jiyu Cheng, Lin Zhang, and Wei Zhang% <-this % stops a space
\thanks{Jinfan Zhou, Jiyu Cheng, Lin Zhang are with the School of Control Science and Engineering, Shandong University, Jinan 250061, China (e-mail:{\tt\small zjf@mail.sdu.edu.cn}; {\tt\small jycheng@sdu.edu.cn}; {\tt\small zl935546110@gmail.com}).}%

\thanks{Wei Zhang is with the School of Control Science and Engineering, Shandong University, Jinan 250061, China, and also with the Institute of Brain and Brain-Inspired Science, Shandong University, Jinan 250061, China (e-mail:{\tt\small davidzhang@sdu.edu.cn}).}%

\thanks{This work was supported in part by the National Key Research and Development Project of New Generation Artificial Intelligence of China under Grant 2018AAA0102504 ; in part by the National Natural Science Foundation of China under Grant U1913204, and Grant 61991411; and in part by the Shandong Major Scientific and Technological Innovation Project (MSTIP) under Grant 2019JZZY010428 and in part by the Fundamental Research Funds of Shandong University under Grant 31400061340342.}

}

\begin{document}

\maketitle
\thispagestyle{empty}
\pagestyle{empty}

%%%%%%%%%%%%%%%%%%%%%%%%%%%%%%%%%%%%%%%%%%%%%%%%%%%%%%%%%%%%%%%%%%%%%%%%%%%%%%%%
\begin{abstract}
We aim to improve the performance of multi-agent flocking behavior by quantifying the structural significance of each agent. We designed a confidence score(ConfScore) to measure the spatial significance of each agent. The score will be used by an auxiliary controller to refine the velocity of agents. The agents will be enforced to follow the motion of the leader agents whose ConfScores are high. We demonstrate the efficacy of the auxiliary controller by applying it to several existing algorithms including learning-based and non-learning-based methods. Furthermore, we examined how the auxiliary controller can help improve the performance under different settings of communication radius, number of agents and maximum initial velocity.
\end{abstract}

% \begin{IEEEkeywords}
% Robot Swarms, Multi-agent System, Flocking
% \end{IEEEkeywords}

\section{Introduction}
Multiple collaborative agents can form large scale swarms which have wide application in various fields including assisting public safety communications~\cite{merwaday2015uav},  environment mapping and exploration~\cite{rencken1993concurrent, burgard2000collaborative},  and cooperative hunting~\cite{cao2006cooperative}. A centralized controller is capable of handling these problems when the scale of the swarm is moderate, while it fails as the scale becomes much larger. Thus, a decentralized  method is an appropriate solution to deal with this situation. In a decentralized system, each agent makes its own decision based on the local information it collects by itself and the information shared by its neighbours. 
% And it would be encouraging to have a generic tool to help improve the performance of decentralized methods.

% We introduce a simple and local method to evaluate the structural significance in its neighbourhood and then use it to produce an assistant acceleration to adjust the velocity of each agent.  

Flocking is a task to coordinate the motion of several autonomous agents which is closely related to the natural animal behaviours~\cite{vicsek2001question, low2000following}. The early research in ecology and biology~\cite{breder1954equations, warburton1991tendency, okubo1986dynamical} has inspired the multi-agent flocking research. A computer animation program~\cite{reynolds1987flocks} mimicking the animal aggregation behaviour was proposed and paved the path for several following work leading to the creation of a new research area called artificial life in computer graphics~\cite{terzopoulos1999artificial}. Flocking is also an important problem in control and robotics~\cite{vasarhelyi2018optimized}, and has many practical applications especially in the control of UAVs~\cite{vasarhelyi2014outdoor, schilling2019learning,hu2020vgai}

Some previous work uses machine learning methods like imitation learning  ~\cite{li2021message,ross2010efficient, ross2011no} and reinforcement learning~\cite{khan2020graph} which require a lot of time to train the controller. While other methods like ~\cite{tanner2004flocking,reynolds1987flocks, jadbabaie2003coordination, tanner2003stable} require no training but will be very difficult to find an optimal controller in a distributed setting.

As proposed in ~\cite{tolstaya2020learning}, graph neural networks~\cite{gama2018convolutional, wu2020comprehensive, kipf2016semi} are suited for the decentralized control system where communication is limited and agents can only share information with their nearby peers. Aggregation graph neural networks are useful when dealing with information sharing, but it aggregates information only depending on the adjacency relationship and ignores the significance of it. Veli{\v{c}}kovi{\'c} \emph{et al.}~\cite{ velivckovic2017graph} combined attention mechanism with graph neural networks but they only considered point-to-point attention and did not use the information provided by neighbours to measure each agent's structural significance in a point-to-group manner. Moreover, such proposed attention can only be used in a learning-based method, which means that it needs extra time to train the network.

 %[5] introduces an attention mechanism to selectively aggregate information shared between agents but it 
 
%  All of the agents in a swarm do not take same role. For example, different from bird or fish flocks, members in a herd only move in a 2-D plane and most of the time the effect of the opinion leaders is very important. It is a decentralized system where multiple opinion leaders take effect together and the formation of these leaders are spontaneous and varying in accordance with the evolution of the dynamic process. The leaders take the major responsibility to maintain the cohesion and motion of the herd.

\begin{figure*}[ht]
\centering
\subfigure{
\includegraphics[width=12cm]{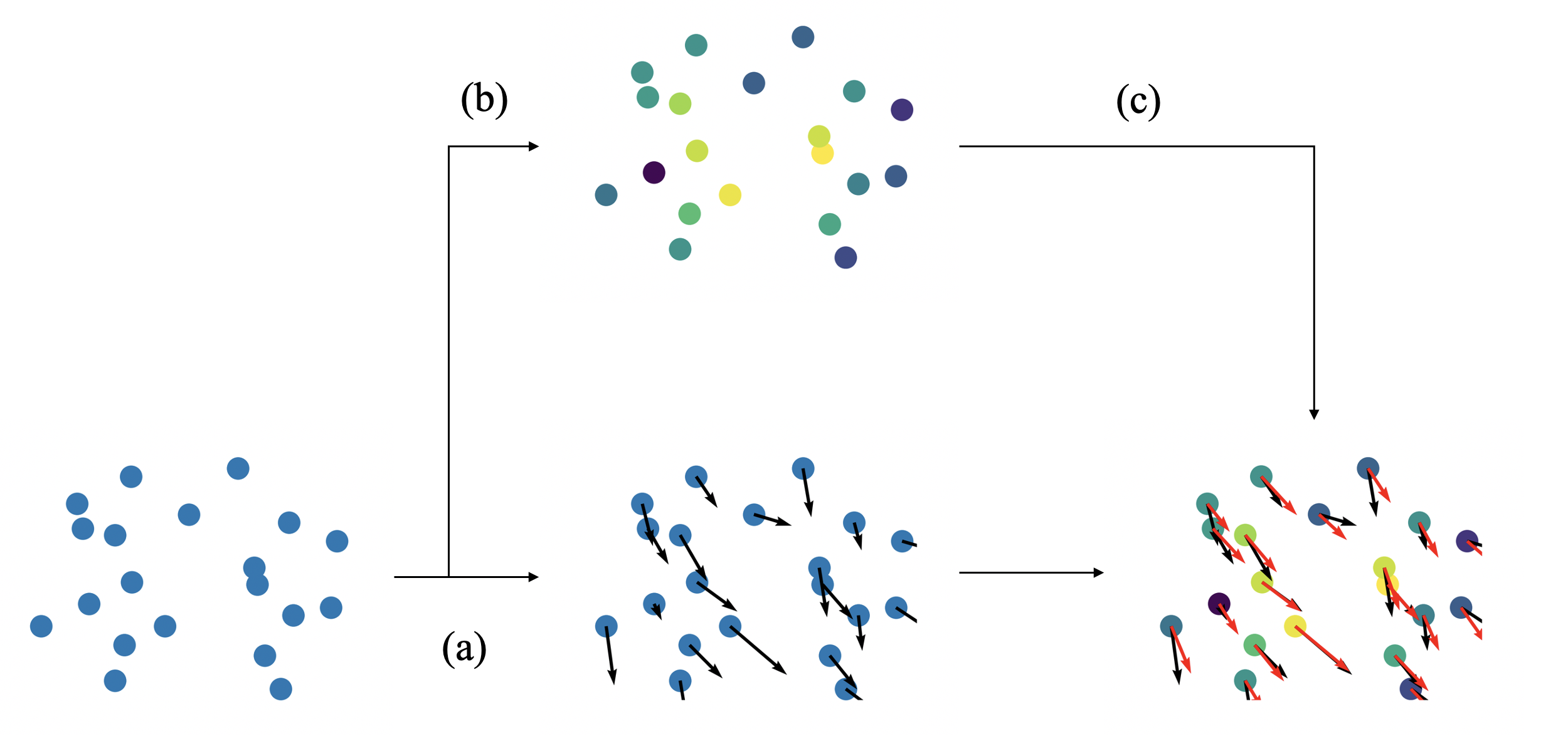}
}
\caption{ConfScore-based auxiliary controller. (a)A centralized/decentralized controller $\Phi$ takes state value $\pmb{x}$ as input and output control action $\pmb{u}$. (b)ConfScores are calculated according to the state value $\pmb{x}$. (c)Based on the ConfScores, we calculate the assistant acceleration $\pmb{\bar{u}}$ to adjust the agents' motion.}
\label{fig::procedure}
\end{figure*}

In order to address these issues, we introduce the ConfScore to comprehensively measure the quality of each agent's position and the consistency of its motion with neighbours. We also propose an auxiliary controller based on the ConfScore for better coordination. We examine the dynamic structural importance of each agent and enforce the agents to follow those opinion leaders in a swarm. It proves to be beneficial to a wide variety of algorithms and requires no extra training. The whole procedure is illustrated as in Fig. \ref{fig::procedure}.

%===============================================================================

%===============================================================================

\section{Control of Multi-agent Swarm}
\label{sec:controlof}

We consider there are $N$ agents in a swarm involved in some dynamic process which requires to be controlled. The process can be characterized by some state value $ \pmb{x}_{i}(t) \in \mathbb{R}^{p} $ 
which demonstrates the states including position, velocity and acceleration of each agent in time $t$. Also, some control action  $\pmb{u}_{i}(t)\in \mathbb{R}^{q}$ is also needed so as to illustrate the action that agents take so as to realize the goal of controlling the whole system.  We denote matrix $\pmb{x}(t)=[\pmb{x}_{1}^{T}{(t)};... ;\pmb{x}_{N}^{T}(t)] \in \mathbb{R}^{N\times p}$ as a collection of the states of all of the agents in a swarm and the matrix $\pmb{u}(t)=[\pmb{u}_{1}^{T}(t);... ;\pmb{u}_{N}^{T}(t)] \in \mathbb{R}^{N\times q}$  as a collection of the control action of the system in time $t$.  The evolution of the dynamic process can be formulated as a differential equation: $\dot{\pmb{x}}(t) = f(\pmb{x}(t),\pmb{u}(t) )$.

Our controller operates in discrete time with a sampling time $T_S$ and the index $n$.  During one sampling time, the control action $\pmb{u}(n)$ is kept the same from time $nT_s$ until $(n+1)T_s$. As the state value matrix $\pmb{x}_n = \pmb{x}(nT_s)$, we can accumulate $\dot{\pmb{x}}(t)$ from time $nT_s$ to $(n+1)T_s$  and give a formulation of the discrete dynamic system as:

\begin{equation}
\pmb{x}_{n+1}=\int_{nT_s}^{(n+1)T_x}f(\pmb{x}(t),\pmb{u}(t))dt + \pmb{x}_n  %\tag{1}
\end{equation}

Based on different tasks we are dealing with, we have different cost function $l(\pmb{x}(t),\pmb{u}(t) )$. So at each time step $nT_s$, we have an immediate cost $l_n = l(\pmb{x}(n),\pmb{u}(n) )$. The objective of the control policy is to minimize the accumulative cost $\sum_{n=0}^{\infty}l_n$. 

% As for a centralized controller, a collection of all the state values of the system $\pmb{x}(n)=[\pmb{x}_{1n}^T;... ;\pmb{x}_{Nn}^T] \in \mathbb{R}^{N\times p}$ is accessible and a policy $\pmb{\pi}$ handling with global information can be formed to 

% A controller should choose control actions $\pmb{u}_n=\pmb{\pi}(\pmb{x_n})$ so as to minimize the cost function. If a centralized controller has access to process dynamics $f(\pmb{x}(t),\pmb{u}(t) )$, then the policy can be readily optimized via several optimization methods~\cite{zhou1996robust, bemporad2002explicit} to get an optimal policy $\pmb{\pi}^*$. In this paper, we mainly focus on a decentralized method that only have access to local information to take actions and can help improve existent decentralized methods together with centralized methods.

We use a graph $\mathcal{G}$ to denote an agent network and use $\pmb{n}$ to represent the collection of all the agents. we define a communication radius \textbf{$R$} . If the distance between two agents  $\pmb{n}_i$ and  $ \pmb{n}_j$are within \textbf{$R$} , then we add an edge $(i, j)\in \mathcal{E}_{n}$ to the graph and   $\pmb{n}_i$ and  $ \pmb{n}_j$ are treated as neighbours. Thus we can define the neighbourhood of $\pmb{n}_i$ at time $n$ as $\mathcal{N}_{ij}=\lbrace j:(i,j)\in \mathcal{E}_{n} \rbrace$. 

% For each agent $\pmb{n}_i$, it can access the information provided by all its neighbours in the neighbourhood and in turn it can dispatch its own information to its neighbours. We utilize this two-way information sharing to (1) measure the the motion quality of each agent via the information it gathered from its neighbours and (2) transmit the information containing importance to its neighbours to help make better decision.

%===============================================================================

\section{ConfScore-based Auxiliary Controller }
\label{sec:methods}

\begin{algorithm}
    \caption{Computation of Assistant Acceleration Based on ConfScore}
    \label{alg::acceleration}
    \begin{algorithmic}[1]
        \Require
            \pmb{$n$}: collection of agents;
            \pmb{$\mathcal{N}$}: collection of neighbourhood of all agents;
            \pmb{$v$}: collection of velocity of all agents;
            $\lambda$: a heuristic magnitude coefficient
        \Ensure
            \pmb{$c$}: collection of ConfScores of all agents
            $\bar{\pmb{u}}$: collection of assistant acceleration of all agents
        \State Initialize \pmb{$c$}, $\bar{\pmb{u}}$ to be \pmb{0}
        \For{each agent $\pmb{n}_{i}$ in \pmb{$n$}}
            \For{ each neighbour $\pmb{n}_{j}$ in $\pmb{n}_{i}$'s neighbourhood $\mathcal{N}_i$}
                \State $c_{\pmb{n}_{i}} = c_{\pmb{n}_{i}} + \frac{\boldsymbol{v_{\pmb{n}_{i}}}\cdot \boldsymbol{v_{\pmb{n}_{j}}} }{||\boldsymbol{v_{\pmb{n}_{i}}}||\cdot||\boldsymbol{v_{\pmb{n}_{j}}} ||}$
            \EndFor
        \EndFor
        
        \For{each agent $\pmb{n}_{i}$ in \pmb{$n$}}
            \State $counter = 0$
            \For{ each neighbour $\pmb{n}_{j}$ in $\pmb{n}_{i}$'s neighbourhood $\mathcal{N}_i$}
            \If{$c_{\pmb{n}_{j}} > c_{\pmb{n}_{i}}$ \textbf{and} $\pmb{n}_{j}$ is among $\pmb{n}_{j}$'s top-k neighbours with regard to \pmb{$c$}}
                \State $\bar{\pmb{u}}_{\pmb{n}_{i}} = \bar{\pmb{u}}_{\pmb{n}_{i}}+ {\lambda}(C_m-C_i)(\pmb{v_{\pmb{n}_j}}-\pmb{v_{\pmb{n}_i}})$
                \State $counter += 1$
            \EndIf
            \EndFor
            \State $\bar{\pmb{u}}_{\pmb{n}_{i}} = \frac{\bar{\pmb{u}}_{\pmb{n}_{i}}}{counter}$
        \EndFor
        \State \Return{\pmb{$c$}, $\bar{\pmb{u}}$}
    \end{algorithmic}
    
\end{algorithm}

% As different agents take different positions in a swarm, they intrinsically have different structural significance, for example, the abnormal behaviour of a marginal agent may not cause too much harm to the entire swarm, yet that of a central agent who has a lot of neighbours and takes up a centric position may lead to the crush of the whole system. So we first propose a ConfScore to comprehensively measure the importance as well as the motion quality of an agent in the swarm. We evaluate it based on two criteria \textbf{(1)} the number of the neighbours and \textbf{(2)} to which extent its velocity is in accordance with its neighbours. Criteria (1) measures the goodness of an agent's position.If one agent only has few neighbours, then it is likely that the agent has already be apart from the main-troop and vice versa. Criteria (2) is used to decide an agent's motion quality, since it is very easy to understand that an agent should be considered as a good member in a swarm if its velocity accords with its neighbours. Thus we propose a ConfScore $C_i$ to measure the extent to which one agent should be confident about its current motion

As different agents take different positions in a swarm, they intrinsically have different structural significance. We first propose a ConfScore to comprehensively measure the structural importance as well as the motion quality of an agent in the swarm. We evaluate it based on two criteria \textbf{(1)} the number of the neighbours and \textbf{(2)} to which extent its velocity is in accordance with its neighbours. Criterion (1) measures the goodness of an agent's position and criterion (2) is used to decide an agent's motion quality. Thus we propose a ConfScore $C_i$ to measure the extent to which one agent should be confident about its current motion.

\begin{equation}
C_i = \sum_{j\in \mathcal{N}_{i}}\frac{\boldsymbol{v_{i}}\cdot \boldsymbol{v_j} }{||\boldsymbol{v_i}||\cdot||\boldsymbol{v_j} ||} %\tag{9}
\end{equation}

% In a flocking problem, direction matters much more than the magnitude of the velocity since it is obvious that a swarm with all agents heading to different directions yet sharing the same speed will perform much worse than another swarm with all agents heading to the same direction but with different speed. For now, in regard with the consistency of the velocity, we focus more on the direction of the agents. So we use cosine similarity to depict the consistency of the velocity between one agent and its neighbours which will be a scalar falling into the range of \textbf{[-1, 1]} and is an proper quantitative measurement of criteria (2).

% We use cosine similarity to depict the consistency of the velocity between one agent and its neighbours. The result will be a scalar falling into the range of [-1, 1] and is a proper quantitative measurement of criterion (2).

% As for criteria (1), a summation over the cosine similarity between each agent and its neighbours will suffice.
One agent would have higher score if it has more neighbours sharing similar velocity with it and vice versa.

% The ConfScore takes into account the number of neighbours together with the consistency between the velocity of the agent and its neighbours. And the calculation of it is completely local and can be applied to both centralized and decentralized algorithms.

After the computation of the ConfScores, every agent is assigned with one score. The score can be interpreted as how confident an agent should stick to its current motion or change its velocity in order to follow other agents who are more likely to be on the right track. Then we utilize ConfScores to compute assistant acceleration as in Algorithm \ref{alg::acceleration}.

$\lambda$ is a coefficient used to determine the extent to which we wish the agent to follow its leaders. It is set heuristically to be number of agents $\frac{30}{\pmb{N}}$ for a non-learning-based algorithm and $\frac{15}{\pmb{N}}$ for a learning-based algorithm. 
% And it is suggested to set a numerical bound to limit the range of the result for stability. We can then use the auxiliary controller to enforce the agents to follow the leaders in its neighborhood.
% % Specially, if we set $k$ equals to 1, then equation (13) becomes 
% % \begin{equation}
% % \bar{u}={\lambda}(C_m-C_i)(\pmb{v_m}-\pmb{v_i}) %\tag{11}
% % \end{equation}

% so that the agent only needs to consider the only neighbour with the highest score in its neighborhood.

The scalar difference $(C_{m}-C_{i})$ shows the extent to which the agent is enforced to follow the motion of its neighbours. If both agents are similarly confident, their neighbours' motion will not have much influence while as the difference gets larger, the impact of the neighbours will be stronger.

The vector difference $(\pmb{v_m}-\pmb{v_i})$ is used to adjust the velocity. Note that only when the two vector $\pmb{v_m}$ and $\pmb{v_i}$ is completely same both in direction and magnitude will it take no effect, in other words, even if two agents move in same direction, still, the auxiliary controller will force them to be at the same speed.

We can generalize flocking algorithms using some policy $\Phi$ to be 
\begin{equation}
\pmb{u}= \Phi(\pmb{x}) %\tag{7}
\end{equation}

So it will be convenient to modify the final control action by an assistant control action $\bar{\pmb{u}}$ using policy $\bar{\Phi}$ as

\begin{equation} 
    \begin{split}
    \dot{\pmb{u}}&= \Phi(\pmb{x}) + \bar{\Phi}(\pmb{x}) \\
    &= \pmb{u} + \bar{\pmb{u}} 
    \end{split} 
\end{equation}

% The local controller $\pmb{u}_{local}$ in \cite{tanner2003stable} lacks the ability to constrain the agents from dispersing as each of the agent doesn't distinguish its different neighbours and only considers the average velocity of its neighbours together with collision avoidance. This deficiency is caused by the lost or disturbed focus as agents are not sure about whether the neighbour it follows is a good leader or not. On the other hand, 
Our ConfScore gives a good measurement to the agents and the auxiliary controller can push the agent to follow at least one best neighbour, which prevents agents losing communication with other agent .

%===============================================================================

\section{Experiment}
We apply the auxiliary controller to flocking controller proposed in \cite{tanner2003stable, tolstaya2020learning} which cover centralized, decentralized, learning-based and non-learning-based methods to see how the auxiliary controller can improve the performance of different kinds of methods. We examine the performance under different settings of number of agents \pmb{$N$}, communication radius \pmb{$R$} and maximum initial velocity \pmb{$V$}. ConfScore-based auxiliary controller is denoted as SA for simplicity. We use the variance in velocities
\begin{equation}
L = {\frac{1}{N}}\sum_{n=1}^{T}\sum_{j=1}^{N}{\left \Vert \pmb{v}_{j,n}-{\frac{1}{N}}\left[ \sum_{i=1}^{N}\pmb{v}_{i,n} \right] \right \Vert}^{2}
\end{equation}
as our cost function.

\begin{figure}[ht]
\centering
\subfigure[Local controller with SA]{
\includegraphics[width=6.5cm]{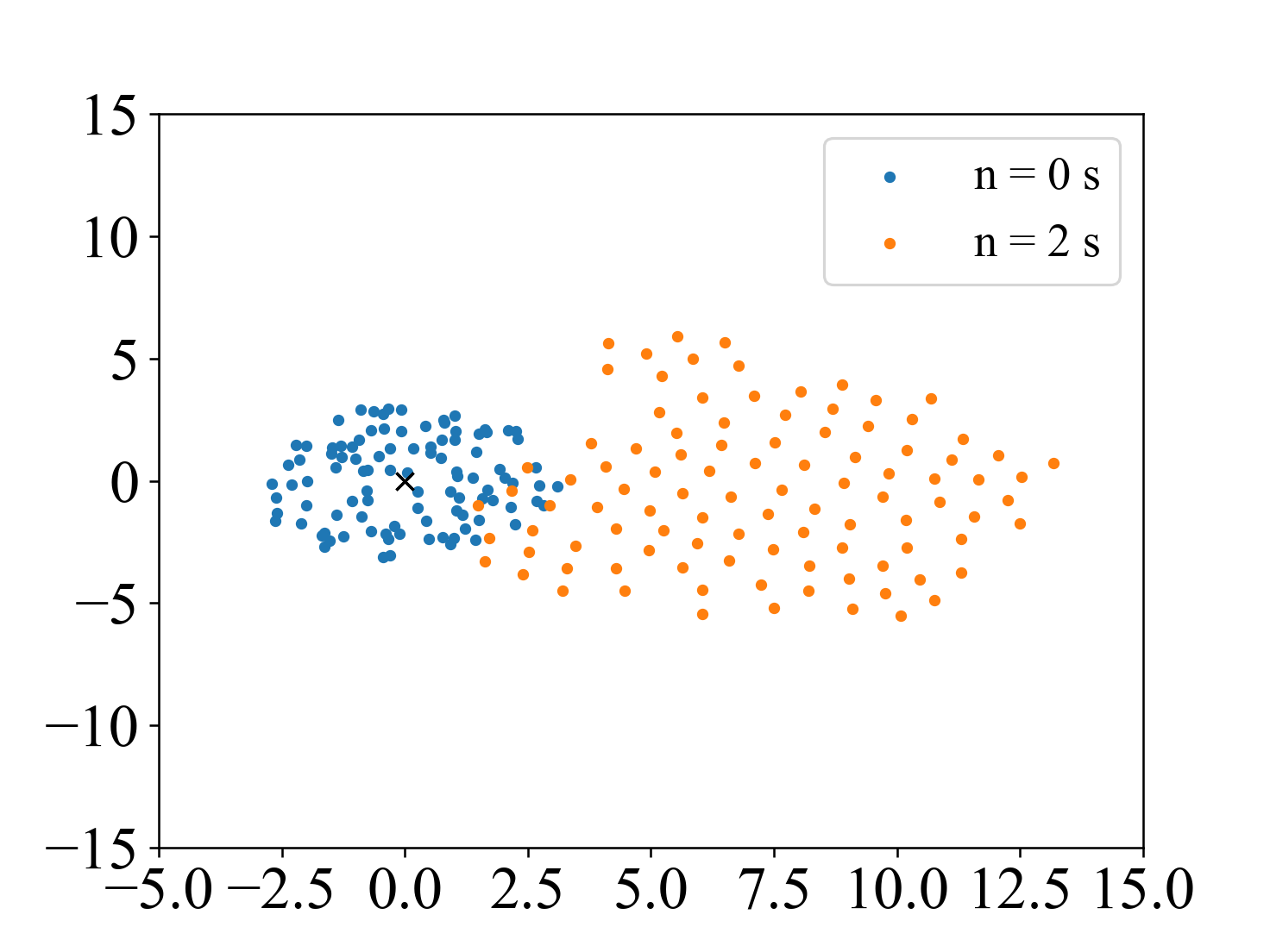}
\label{fig::begin_1}
}
\quad
\subfigure[Local controller]{
\includegraphics[width=6.5cm]{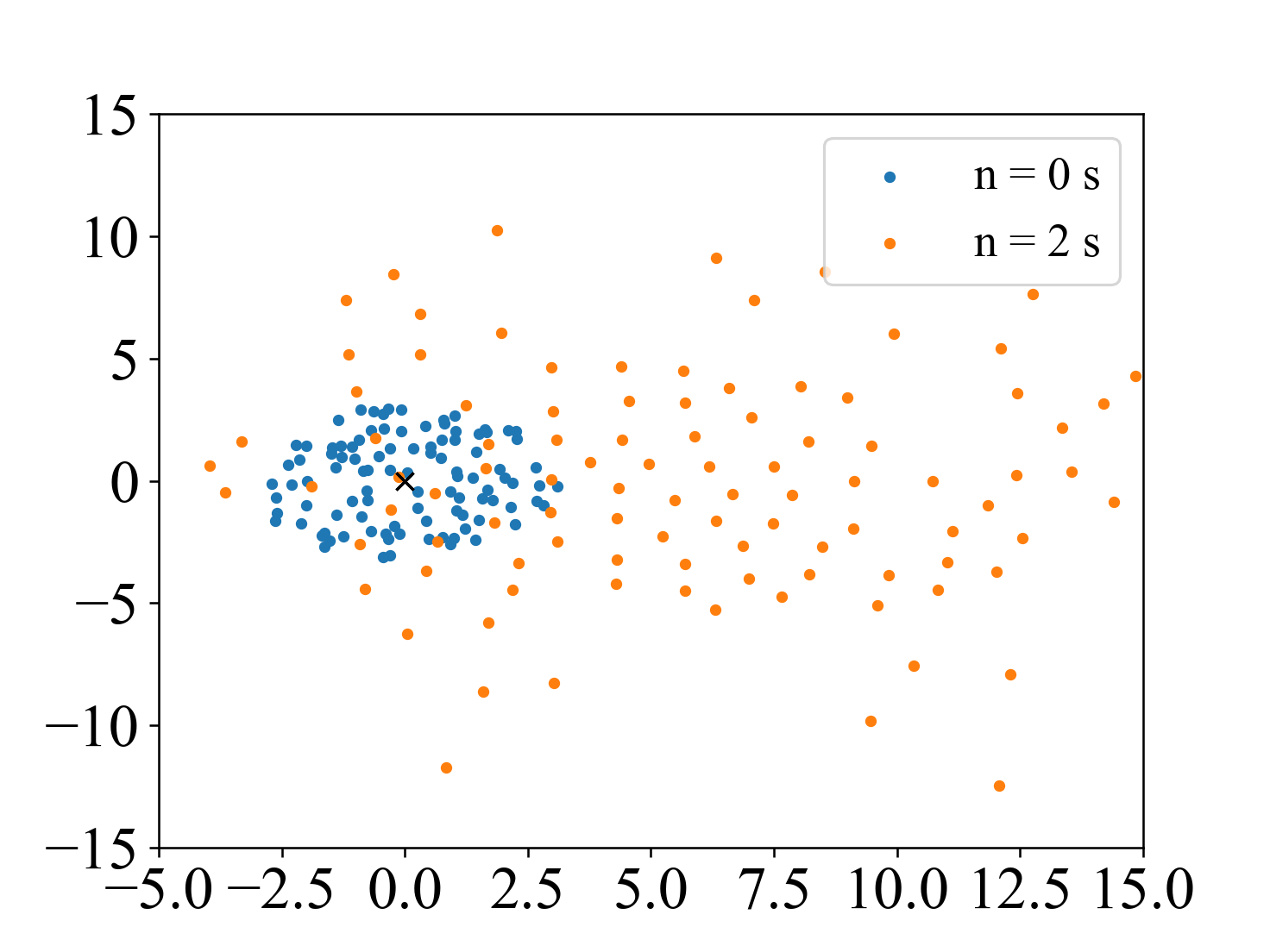}
\label{fig::begin_2}
}
\caption{At the initial state of the process, the auxiliary controller successfully constrains the swarm from scattering, while the local controller cannot. \pmb{$V$} is set to be 3.5m/s, \pmb{$N$}=100 and \pmb{$R$}=1m.}
\label{fig::begin}
\end{figure}

\begin{figure*}
\centering
\subfigure[]{
\includegraphics[width=6.3cm]{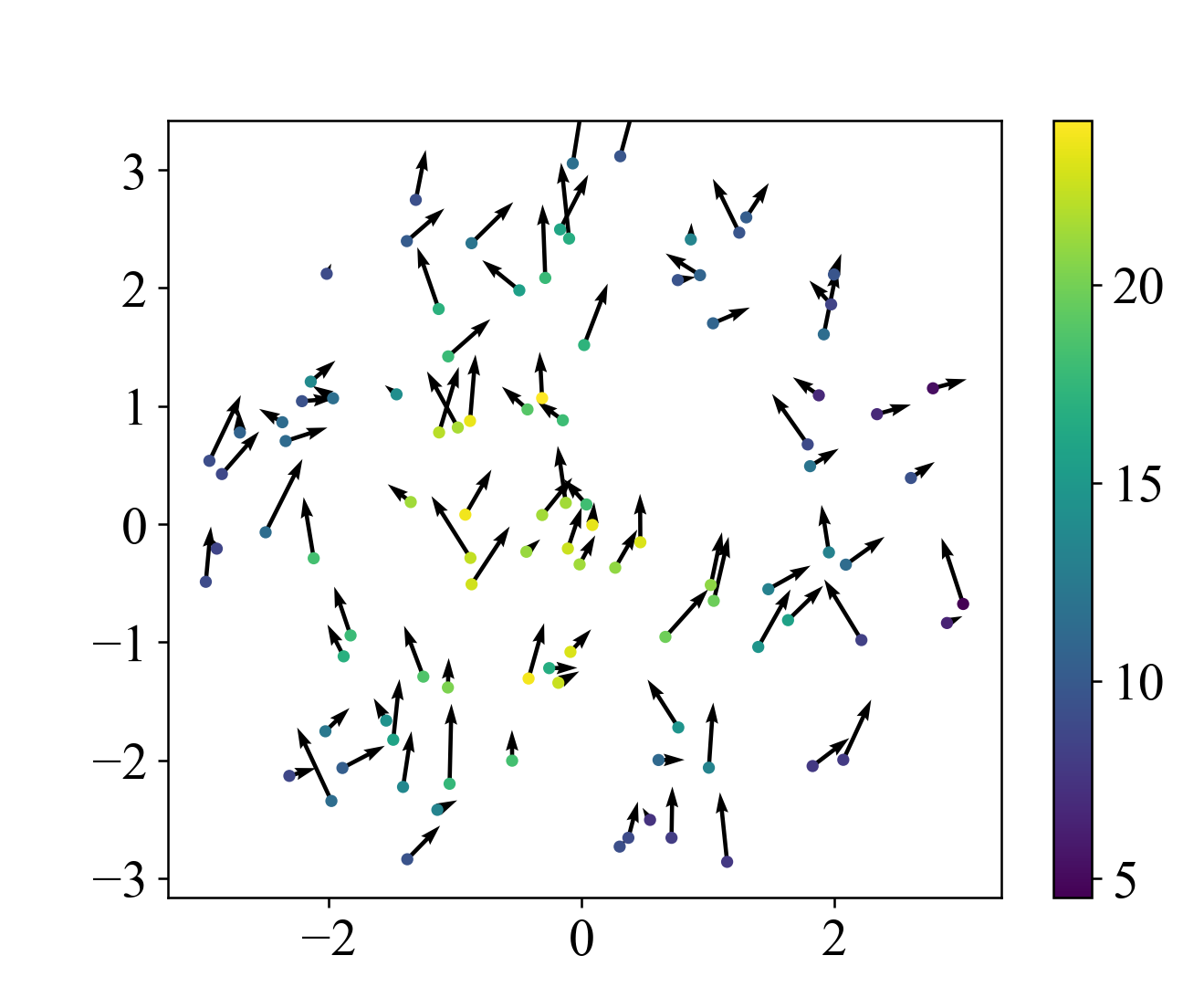}
\label{fig::scores_a}
}
\quad
\subfigure[]{
\includegraphics[width=6.3cm]{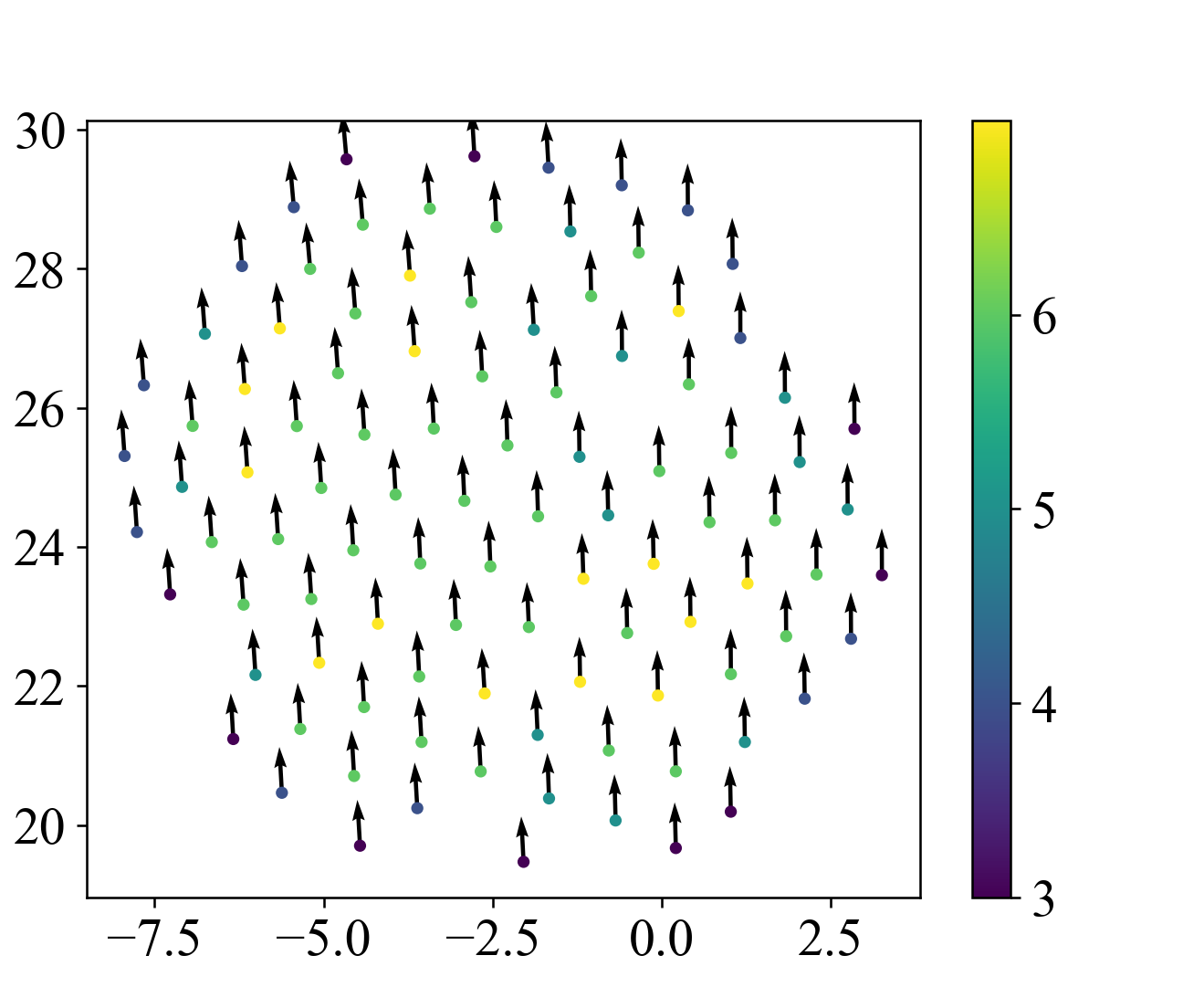}
\label{fig::scores_b}
}
\quad
\subfigure[]{
\includegraphics[width=6.3cm]{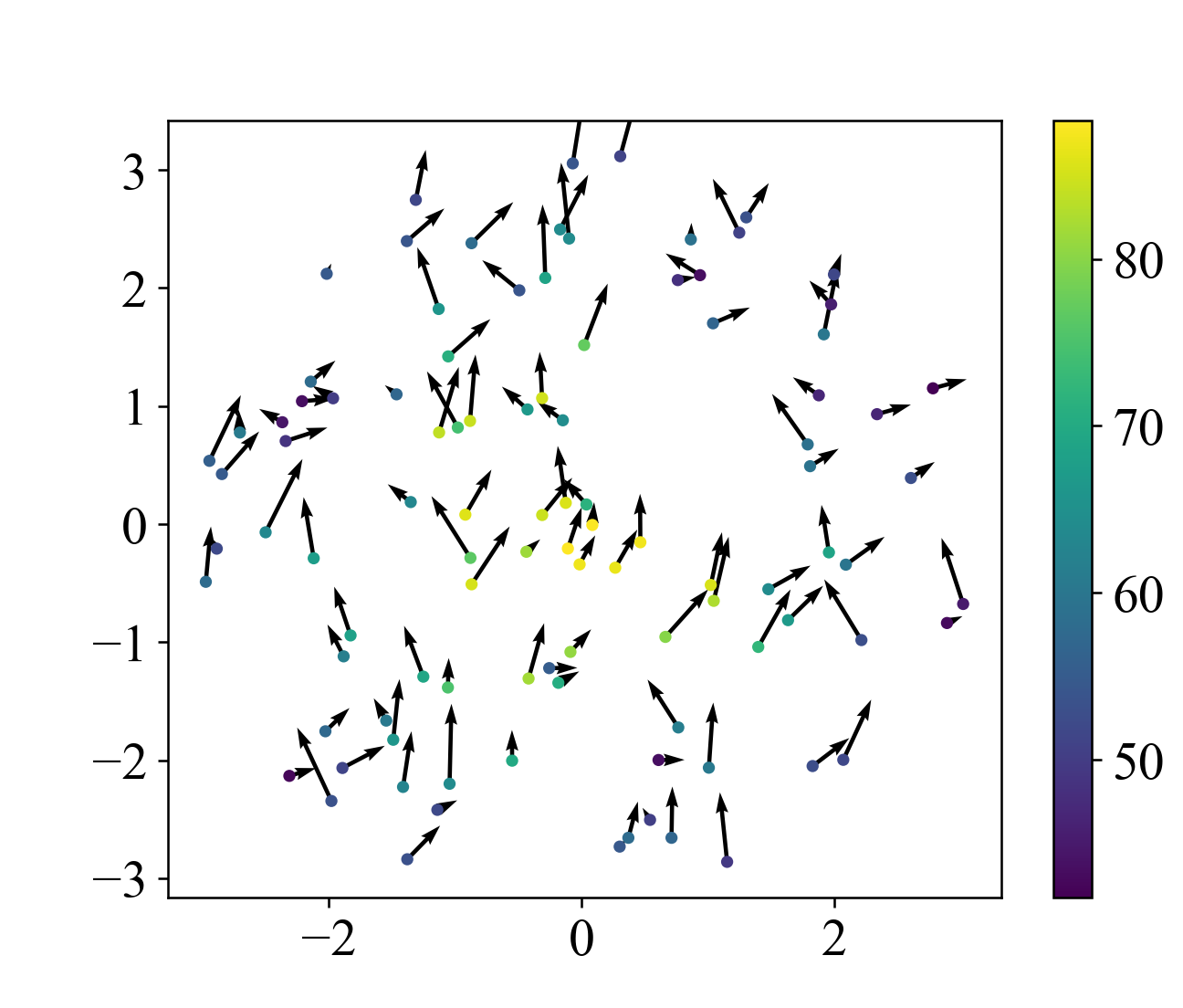}
\label{fig::scores_c}
}
\quad
\subfigure[]{
\includegraphics[width=6.3cm]{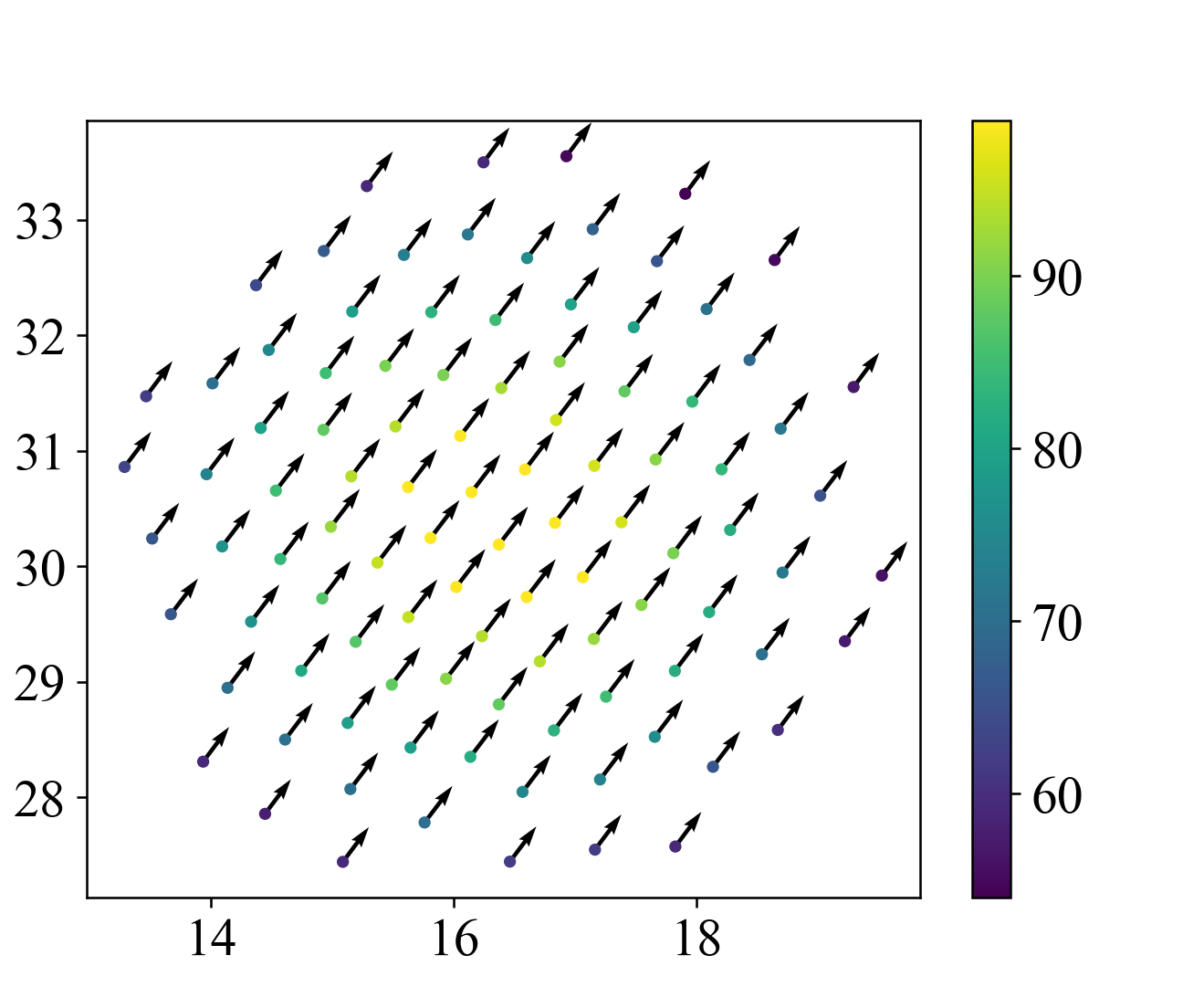}
\label{fig::scores_d}
}

\caption{ConfScores distribution under different communication radius. (a) and (b) show the initial and final scores of the flock under when Comm. Radius is 1.5m while (c) and (d) shows the situation under 4.0m. The cosine similarity is normalized to be in range $[0, 1]$ for better illustration.}
\label{fig::scores}
\end{figure*}

\subsection{Applying to Non-learning-based Method}

$\pmb{u}_{local}$ proposed in \cite{tanner2003stable} can be used as a local controller to make the control action. Its centralized version
\begin{equation}
\pmb{u}_i^*=-\sum_{j\in \pmb{n}} (\pmb{v}_i - \pmb{v}_j) - \sum_{j\in \pmb{n}}\nabla_{r_i}U_{ij} %\tag{5}
\end{equation}
is a more powerful controller yet needs access to the global information. Both $\pmb{u}_{local}$and $\pmb{u}^{*}$ have no learnable parameters. We apply the auxiliary controller to both of the controllers and examine the performance based on different settings. 
For a local controller, the controller should enforce the agents not to disperse sparsely at early stage otherwise the drastic lost of neighbours will quickly occur. The performance of a local controller depends highly on the number of the neighbours each agent have because it relies only on local information to decide its control action. Once an agent completely loses communication with its peers, it will simply keep moving in its original velocity unless it comes into its peers again, which rarely happens.

The auxiliary controller helps improve the robustness of the controller in regard with the maximum initial velocity. As Fig.\ref{fig::begin} shows, the main reason for the local controller to behave poorly as the maximum initial velocity increases is that high velocity will cause the agents to quickly scatter at the very beginning, and it will result in a random motion behaviour. However, the auxiliary controller  can help alleviate this situation in that the magnitude of the auxiliary controller is in proportion to the velocity of the agents and the ConfScore can selectively amplify or shrink the magnitude. So the swarm can keep cohesive even if it is driven by a high velocity. In Fig.\ref{fig::begin_1}, the swarm is driven by the fast initial speed and get scattered. But in Fig.\ref{fig::begin_2}, the agents are still cohesive. We fix the \pmb{$N$} to be 100 and \pmb{$R$} to be $1m$. From Fig.\ref{fig::begin_2} we can see that the performance of a local controller with auxiliary controller is far more stable than a vanilla one.

% As is shown in fig.\ref{fig::data_c}, The power of the assistant acceleration is scalable as the number of the agents increases. Since the computation of the assistant acceleration is completely local, the local connection relationship between agents remains unchanged as the scale of the entire swarm increases.

The number of neighbours will increase as the communication radius increases, thus each agent can sample more neighbours' speed so as to choose a better leader. As is shown in Fig. \ref{fig::scores}, when the swarm is first initialized, due to the randomness of velocity, the distribution of the ConfScores is largely dependent on the position of the agents. Generally, an agent's ConfScore is higher if it is located near the center of the swarm , as it will have more neighbours, and lower if it is closed to the margin. This can cause an phenomenon that the outer agents tend to follow the motion of the inner ones which can help prevent the agents from scattering. As the process proceeds, the distribution of the ConfScores will be attributed to the communication radius \textbf{$R$}. As in Fig.\ref{fig::scores_b} a small \textbf{$R=1.5m$} will result in that the agents with high scores scatter over the entire swarm while in Fig.\ref{fig::scores_d} a large \textbf{$R=4m$} will then make the confident agents gather in the center of the swarm. In consequence, the entire swarm tends to split up if the communication radius is too small while much more robust as the radius increases. 

We also examine different $k$ values. As $k$ increases from 1, instead of only focusing on the leader of the highest score, each agent pays more attention to other neighbours. However, different to simple averaging over the velocity of an agent's neighbours, the assistant acceleration actually only focuses on those neighbours whose scores are higher ignoring those with scores lower than the agent itself, even if they are among its top-k neighbours. Increase in k can enhance the robustness of the acceleration since it is intuitive that the weighted average over the top-k neighbours can avoid the situation where a \textit{fake} leader, whose motion may be in consistency with its neighbours while different from the whole swarm, may be too confident about itself and just split the original swarm leading some agents to move in a wrong way with it.
% As shown in Fig.\ref{fig::data_a}, \ref{fig::data_c}, \ref{fig::data_e}, increase in $k$ can have a little but indeed limited gain, so in general we typically choose $k$ to be 1.

\begin{figure*}
\centering
\subfigure[]{
\includegraphics[width=6.3cm]{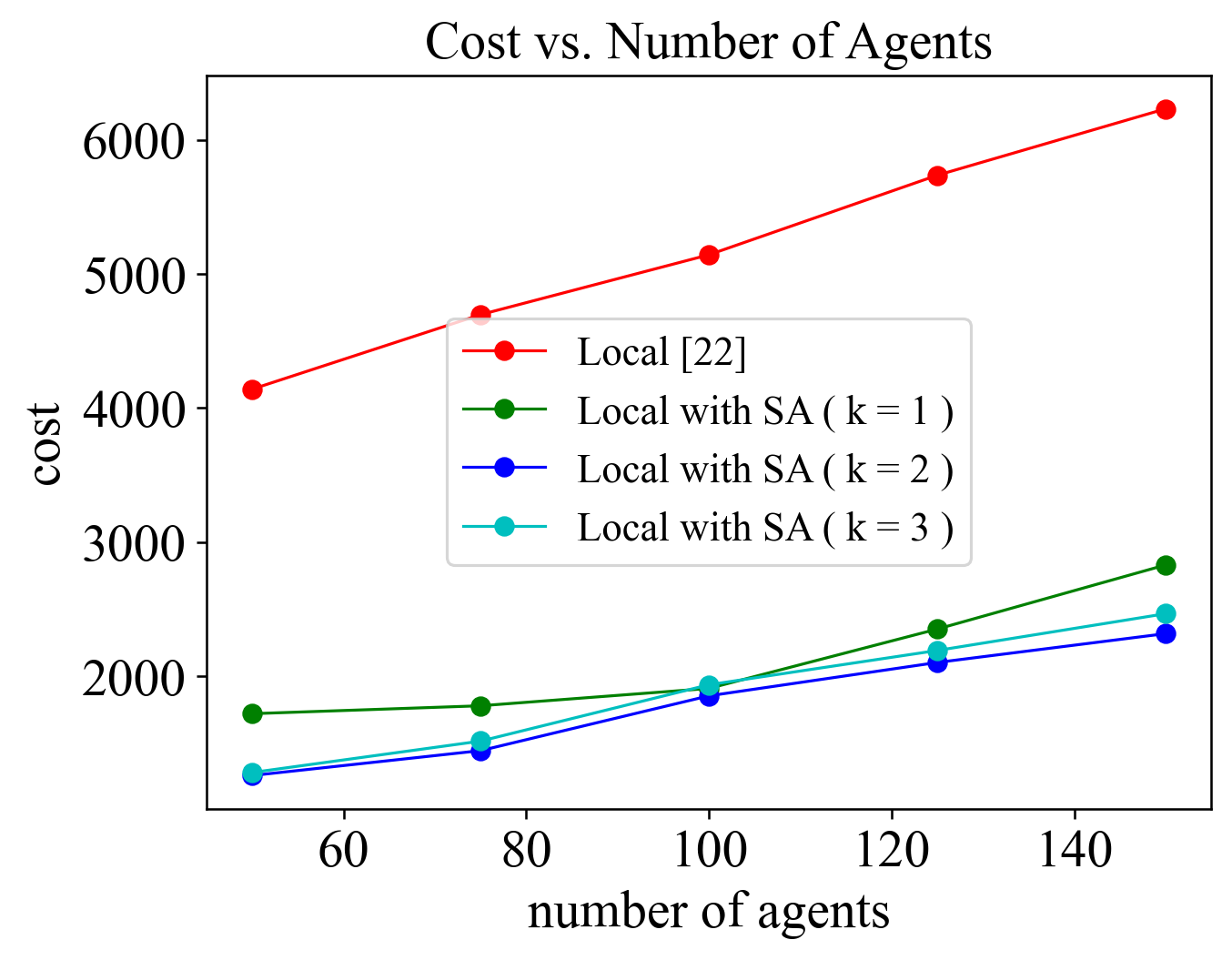}
\label{fig::data_a}
%\caption{fig1}
}
\quad
\subfigure[]{
\includegraphics[width=6.3cm]{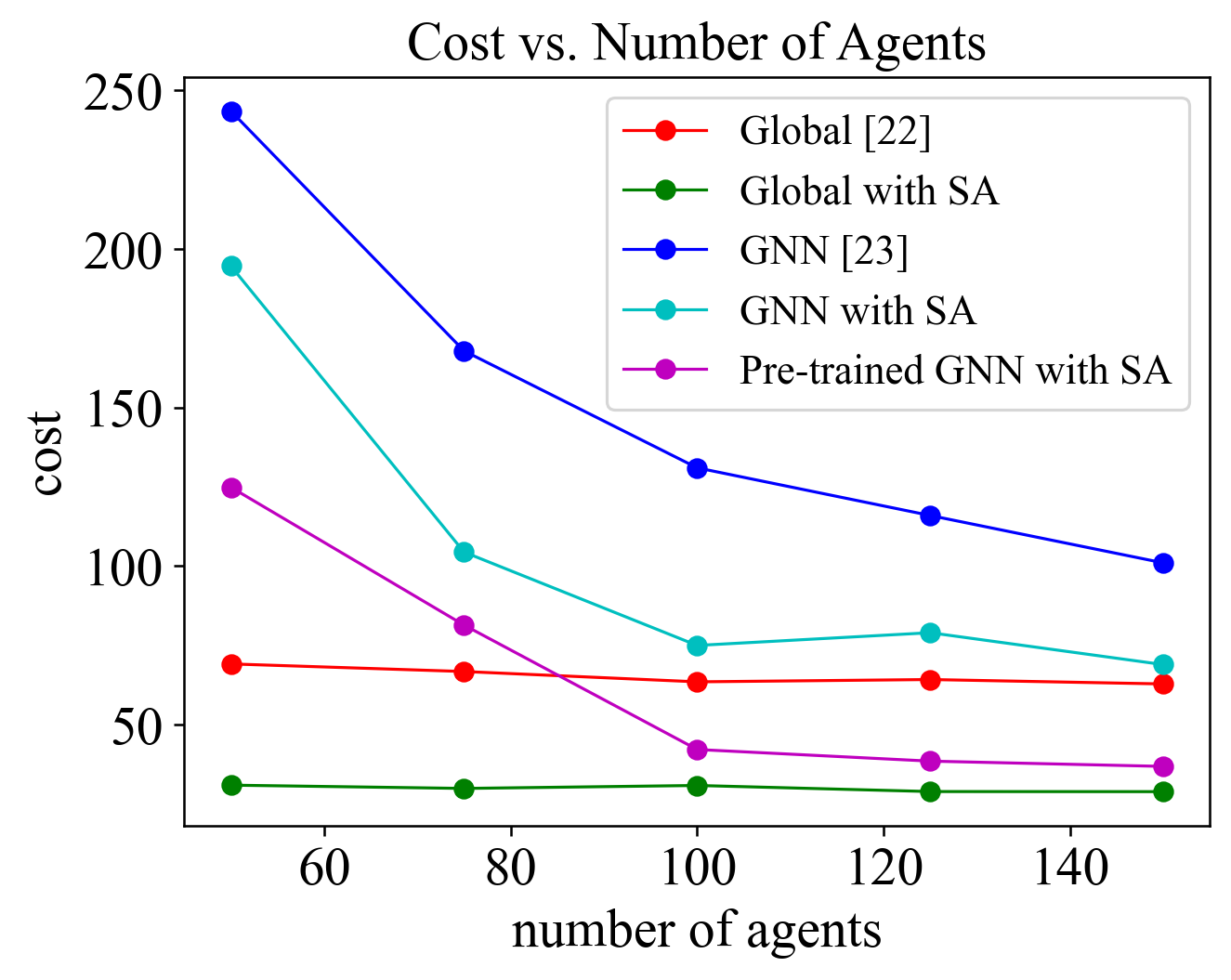}
\label{fig::data_b}
}
\quad
\subfigure[]{
\includegraphics[width=6.3cm]{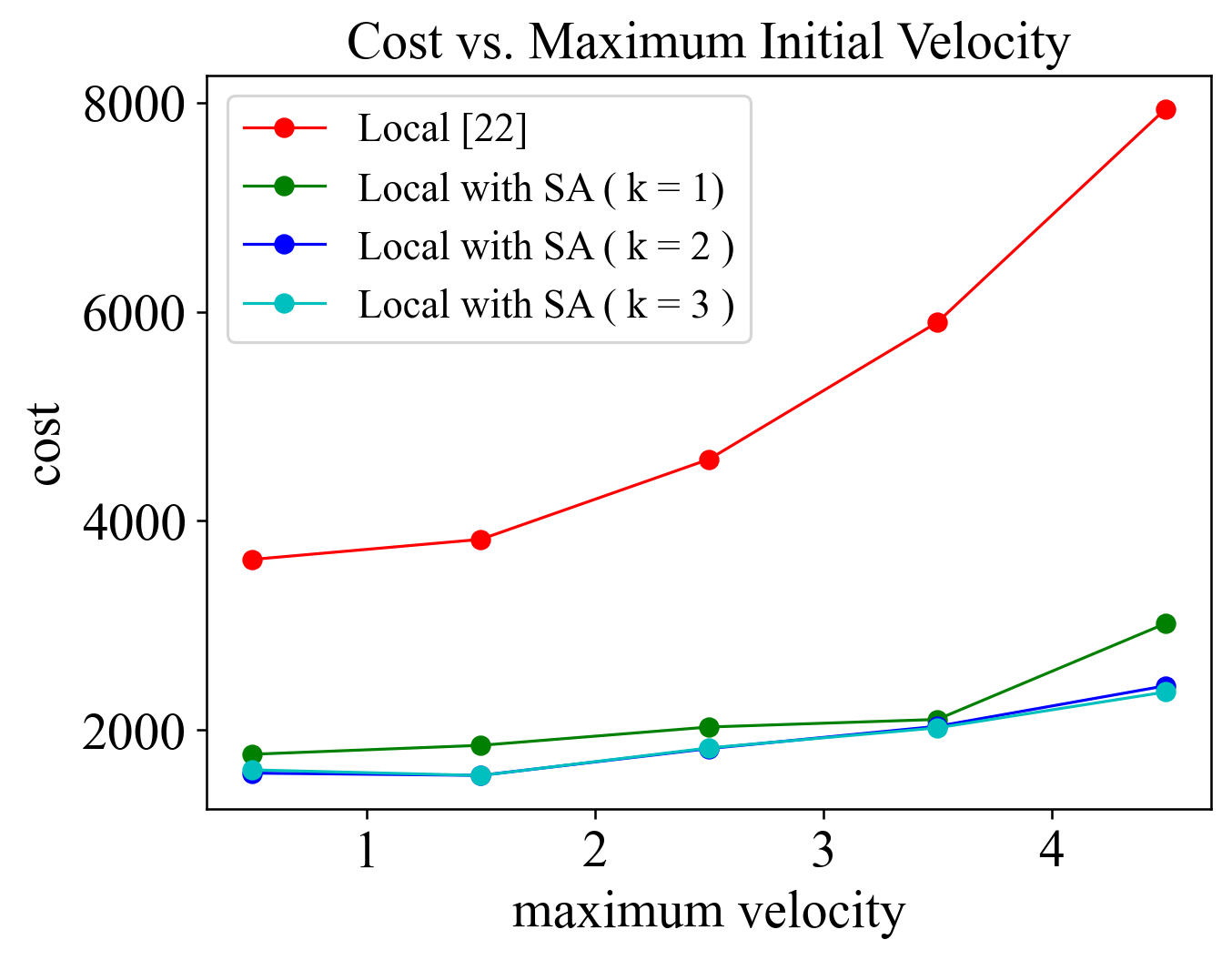}
\label{fig::data_c}
}
\quad
\subfigure[]{
\includegraphics[width=6.3cm]{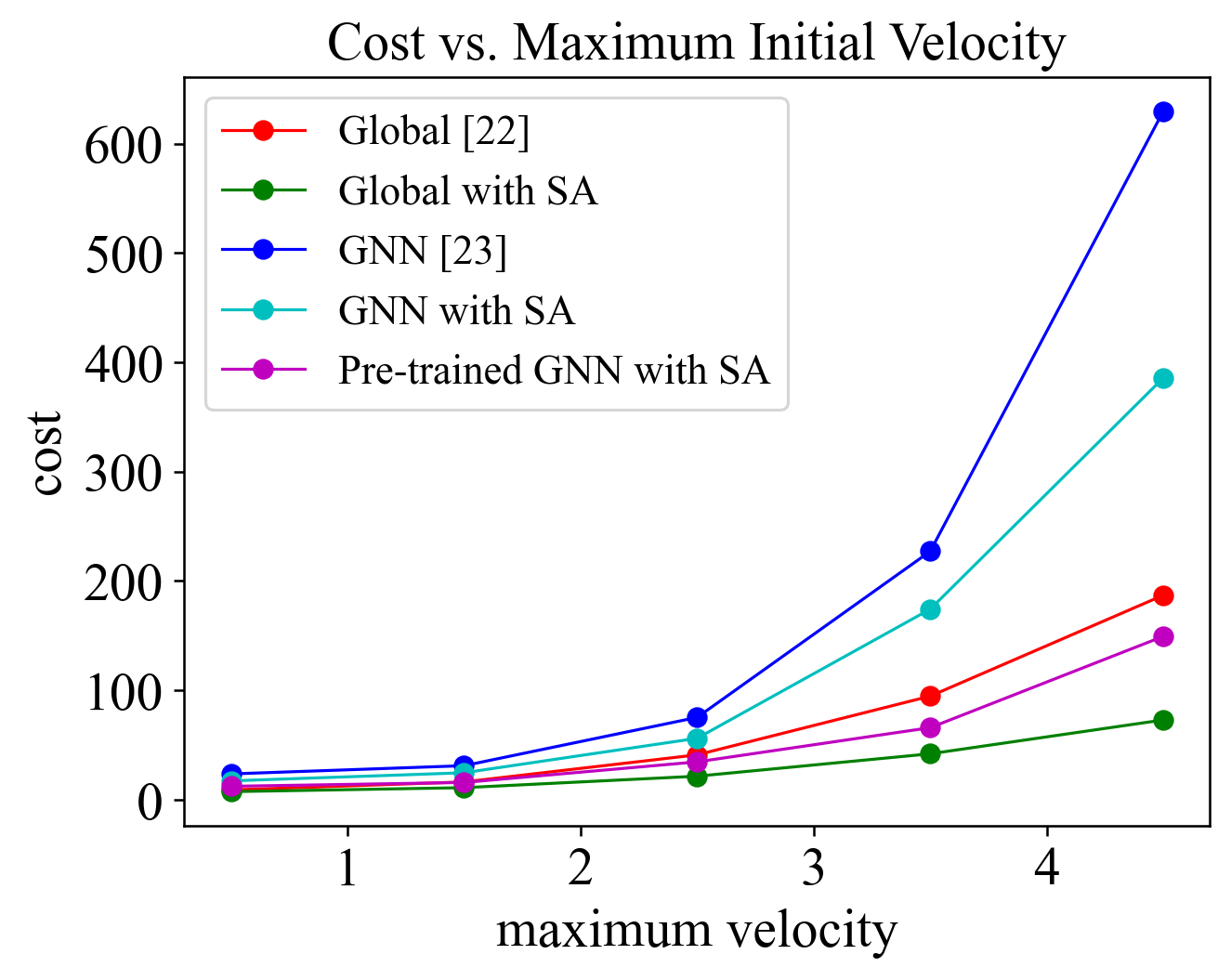}
\label{fig::data_d}
}
\quad
\subfigure[]{
\includegraphics[width=6.3cm]{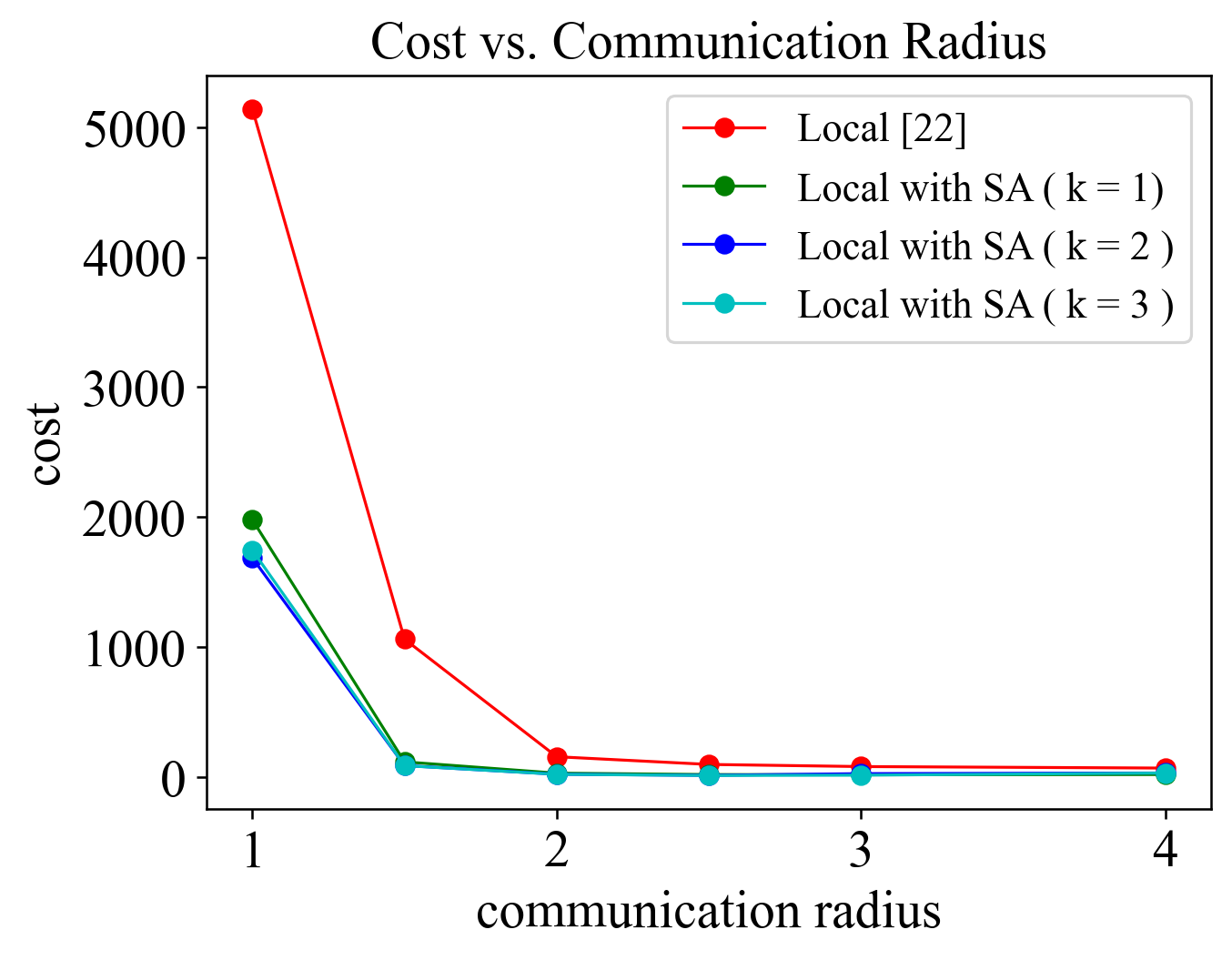}
\label{fig::data_e}
}
\quad
\subfigure[]{
\includegraphics[width=6.3cm]{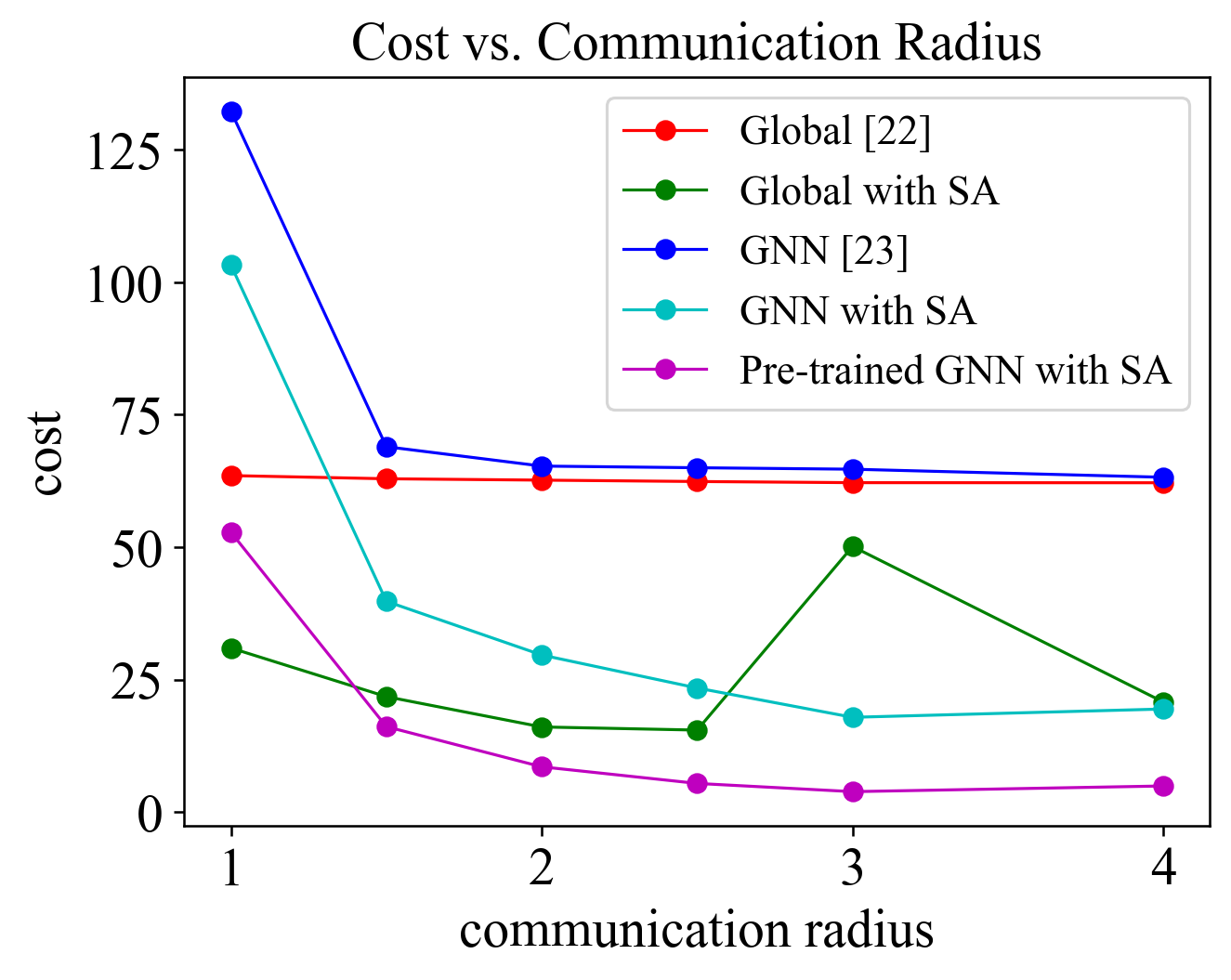}
\label{fig::data_f}
}
\caption{Experimental results on different numbers of agents, maximum initial velocity and communication radius.}
\label{fig::data}
\end{figure*}

We also test on the global controller $\pmb{u}_{*}$ to see how it works. As we can see from fig.\ref{fig::data_b}, \ref{fig::data_d}, \ref{fig::data_e}, the auxiliary controller can also help a global controller to achieve better performance, although as the communication radius varies, the controller will sometimes become unstable. In general, still, it is of much benefit.

\subsection{Applying to Learning-based Method}

As is illustrated, the auxiliary controller can be conducive to non-learning methods, likewise, it can be readily applied to a learning-based method. 

DAGNN proposed by \cite{tolstaya2020learning} uses imitation learning to train an aggregation graph neural network as a local controller. It imitates the behavior of a global expert algorithm by utilizing local observation shared by multi-hop information exchange. 

% We train the DAGNN together with assistant acceleration and fix the hop number \pmb{$K$} and the number of hidden neurons to be 2 and 32. It can be seen from fig.\ref{fig::convergence} that the assistant acceleration works as a powerful regulator that at the very beginning, when the parameters of the graph neural network were just initialized, naturally the control action taken by the controller tends to be random,  but the assistant acceleration takes effect independent of the GNN controller in that it has no learnable parameter and thus the performance of DAGNN with the assistant acceleration is largely improved at beginning and further benefits the following training stage. 

% \begin{figure}
% \centering
% \subfigure[]{
% \includegraphics[width=7.0cm]{pics/convergence.png}
% }
% \caption{Cost with regard to the training iteration}
% \label{fig::convergence}
% \end{figure}

The auxiliary controller is compatible with an imitation learning algorithm in that it doesn't cause the failure of convergence. We conduct experiments on different stages when we apply the auxiliary controller to DAGNN, namely we (1) train a DAGNN with auxiliary controller and (2) apply the controller to the output of a pre-trained DAGNN model. It can be seen from Fig.\ref{fig::data_b}, \ref{fig::data_d}, \ref{fig::data_e} the auxiliary controller can largely help the improvement of the DAGNN model in both ways. 
% However, it suggests that it should be better to apply the auxiliary controller to a pre-trained model since the learned controller may not have the characteristic of the auxiliary controller and the application of it during the training phase may disturb the learning process.

Use of the auxiliary controller helps the local DAGNN controller to excel the performance of a global control. The drawback of a local controller mainly lies in the lost of communication caused by scattering and since the auxiliary controller can largely help alleviate the scattering problem, it is of much benefit to the existent local controllers. Further, As the communication radius expands, the performance of a vanilla DAGNN plateaus, the application of auxiliary controller helps the model break through the bottleneck to achieve better performance as shown in Fig.\ref{fig::data_f}.

%===============================================================================
\section{Conclusion}
We have demonstrated that the utility of auxiliary controller is convenient and compatible to various kinds of existing flocking algorithms. We test it under different settings of the scale, communication radius and maximum initial velocity of swarms. We show that the auxiliary controller is adaptive to the scale of the swarm and can improve other algorithms' robustness. It was also discussed that how the distribution of ConfScore varies due to different communication radius and how it takes affect to keep the agents cohesive. We propose the ConfScore to be a proper measurement of an agent's motion quality and the auxiliary controller to be a 
general tool to improve the performance of the flocking task.

%\section*{Acknowledgment}

% \section*{References}

% Please number citations consecutively within brackets \cite{b1}. The 
% sentence punctuation follows the bracket \cite{b2}. Refer simply to the reference 
% number, as in \cite{b3}---do not use ``Ref. \cite{b3}'' or ``reference \cite{b3}'' except at 
% the beginning of a sentence: ``Reference \cite{b3} was the first $\ldots$''

% Number footnotes separately in superscripts. Place the actual footnote at 
% the bottom of the column in which it was cited. Do not put footnotes in the 
% abstract or reference list. Use letters for table footnotes.

% Unless there are six authors or more give all authors' names; do not use 
% ``et al.''. Papers that have not been published, even if they have been 
% submitted for publication, should be cited as ``unpublished'' \cite{b4}. Papers 
% that have been accepted for publication should be cited as ``in press'' \cite{b5}. 
% Capitalize only the first word in a paper title, except for proper nouns and 
% element symbols.

% For papers published in translation journals, please give the English 
% citation first, followed by the original foreign-language citation \cite{b6}.

\printbibliography %Prints bibliography

% \vspace{12pt}
% \color{red}
% IEEE conference templates contain guidance text for composing and formatting conference papers. Please ensure that all template text is removed from your conference paper prior to submission to the conference. Failure to remove the template text from your paper may result in your paper not being published.

\end{document}